\documentclass[12pt]{article} \input epsf.tex
\usepackage{graphicx}
\usepackage{amsmath}
\usepackage{amsfonts}
\usepackage{latexsym}
\usepackage{stmaryrd}
\usepackage{amssymb}

\usepackage{mcite}

\newcommand{\be}{\begin{equation}}
\newcommand{\ee}{\end{equation}}
\newcommand{\bea}{\begin{eqnarray}}
\newcommand{\eea}{\end{eqnarray}}
\newcommand{\bear}{\begin{eqnarray}}
\newcommand{\eear}{\end{eqnarray}}
\newcommand{\RR}{\mathbb R}
\newcommand{\ba}{\begin{array}}
\newcommand{\ea}{\end{array}}

\begin{document}

\pagestyle{empty} \begin{titlepage}
\def\thepage {} % Kill page numbering

\title{\Large \bf Anomaly-Free Sets of Fermions}

\author{\normalsize
\bf \hspace*{-.3cm}
Puneet Batra$^1$,
Bogdan A.~Dobrescu$^2$, David Spivak$^3$
\\ \\
{\small {\it
$^1$ High Energy Physics Division, Argonne National Lab, Argonne, IL 60439, USA}}\\
{\small {\it
$^2$ Theoretical Physics Department, Fermilab, Batavia, IL 60510, USA }}\\
{\small {\it
$^3$ Mathematics Department, University of California, Berkeley, CA 94720, USA}}\\
}

\date{ } \maketitle

\vspace*{-7.4cm}
\noindent \makebox[11cm][l]{\small \hspace*{-.2cm}
hep-ph/0510181}{\small ANL-HEP-PR-05-106}  \\
\makebox[11cm][l]{\small \hspace*{-.2cm} October 13, 2005}{\small 
FERMILAB-Pub-05-451-T} \\

\vspace*{6.9cm}

\begin{abstract}
We present new techniques for finding anomaly-free sets of fermions.
Although the anomaly cancellation conditions typically
include cubic equations with integer variables that cannot be solved
in general, we prove by construction that any chiral set of fermions
can be embedded in a larger set of fermions which is chiral and
anomaly-free.
Applying these techniques to extensions of the Standard Model, we find
anomaly-free models that have arbitrary quark and lepton charges under
an additional $U(1)$ gauge group.

\end{abstract}

\vfill \end{titlepage}

\baselineskip=18pt \pagestyle{plain} \setcounter{page}{1}

%%%%%%%%%%%%%%%%%%%%%%%%%%%%%%%%%%%%%%%%%%%%%%%%%%%%%%%%%%%%%%%%%%%%
\section{Introduction} \setcounter{equation}{0}

Gauge symmetries successfully describe the electromagnetic, weak, and
strong interactions of particle physics. Nevertheless, unitarity and
renormalizability of the Standard Model do not necessarily follow from
a classical invariance of the Lagrangian under $SU(3)_C\times SU(2)_W
\times U(1)_Y$ gauge transformations: one-loop $SU(N)$
and $U(1)$ gauge anomalies must also be absent
\cite{Adler:1969gk,*Bell:1969ts,Bardeen:1969md, Georgi:1972bb,Gross:1972pv}.

In order to avoid these local gauge anomalies, the sum over triangle
diagrams with gauge bosons as external lines and charged fermions
running in the loops must vanish. As a consequence, the sums over loops 
with more external lines and over higher-order diagrams automatically 
vanish \cite{Adler:1969er, Bardeen:1969md}.  Similarly, the sum
over triangle diagrams involving two gravitons and one $U(1)$ gauge
boson in the external lines must vanish, or else this mixed
gravitational-$U(1)$ anomaly will also lead to an explicit breaking of
the gauge symmetry by gravitational interactions (for a review see
\cite{Eguchi:1980jx}).  Finally, the $SU(2)$ gauge symmetry may suffer
from a global gauge anomaly unless the number of Weyl
fermion doublets is even \cite{Witten:1982fp}.  

A remarkable property of all elementary fermions discovered so far is
that they form a chiral set---none of them can have a gauge invariant
mass term.  The cancellation of anomalies within a chiral set of
fermions is highly nontrivial. Given the observed $SU(3)_C \times
SU(2)_W$ representations found in the Standard Model, the anomaly
cancellation conditions are restrictive enough to uniquely determine
the $U(1)_Y$ charges, assuming that not all of them are zero.  
More strikingly, the {\it minimal} anomaly-free chiral set of fermions
charged under $SU(3)_C \times SU(2)_W \times U(1)_Y$ is exactly given
by a Standard Model generation \cite{Geng:1988pr,*Minahan:1989vd,*Geng:1990nh,*Frampton:1993bp}. 
Therefore, anomaly cancellation provides an explanation for the fermion    
structure of the Standard Model which is an alternative to the explanation
provided by grand unified theories, where an entire Standard Model
generation can be embedded in a single, anomaly-free SO(10) 
representation.

Anomaly cancellation will constrain the charges of the
Standard Model fermions under any newly discovered gauge groups,
whether these groups follow from grand unification or not.
Many models of physics beyond the Standard Model
incorporate new gauge groups, and the couplings of the Standard Model
fermions to the gauge bosons associated with these groups are
completely determined by the spectrum of charges. It is therefore
useful to have methods that allow finding, in general, sets of
fermions which are anomaly-free under arbitrary gauge groups. Since
vector-like pairs of fermions do not contribute to these anomalies, a
complete description of anomaly-free sets hinges only on the
identification of all {\it chiral} anomaly-free sets. 

Furthermore, if the chiral set of Standard Model fermions is charged
under a new gauge group, anomaly cancellation usually dictates the
presence of additional fermions. The complete set of fermions under
the full gauge group would most likely be chiral. Imagine instead that the
full theory were completely vector-like---after the breaking of the
extended gauge symmetry to $SU(3)_C \times SU(2)_W \times U(1)_Y$, one
would be left with both the observed Standard Model fields and a
set of conjugate partners. To avoid mixing with the Standard Model
fields, these conjugate partners should have large masses that proceed
through electroweak symmetry breaking, which would induce too 
large corrections to electroweak observables.

In the case of $SU(N)$ gauge groups, comprehensive lists of
anomaly-free sets have been identified using numerical methods
\cite{Eichten:1982pn}.  By contrast, anomaly-free sets of chiral
fermions charged under a $U(1)$ gauge group, or under direct products
of gauge groups including at least one $U(1)$ group, have been less
thoroughly catalogued, despite the common appearance of extra $U(1)$
groups in connection to flavor symmetry (see, {\it e.g.,}
\cite{Ibanez:1994ig, Binetruy:1994ru,*Jain:1994hd,*Dudas:1995yu,*Choi:1996se,
*Irges:1998ax,*Babu:2003zz,*Chankowski:2005qp}),
supersymmetry breaking (see, {\it e.g.,}
\cite{Binetruy:1996uv,*Dvali:1996rj,*Dobrescu:1997qc,*Cheng:1998nb,
*Arkani-Hamed:1998nu}),
neutrino masses \cite{Appelquist:2002mw, Babu:2003is, Leontaris:2004rd,*Kang:2004ix,
Davoudiasl:2005ks,Sayre:2005yh}, and many other model building issues
(see, {\it e.g.,} 
\cite{Cheng:1998hc,*Erler:2000wu,*Demir:2005ti,*Barr:2005je,*Cvetic:1995rj,
*Izawa:1999cm,*Maekawa:2001uk}). Many of
these constructions depend on the existence of a chiral set of fermions.

If the quarks and leptons have arbitrary
charges under a new $U(1)$, then there are a number of gauge anomalies
that need to be cancelled.  Usually, this is achieved by the inclusion
of additional fermions with carefully chosen charges. An alternative
is available in the context of string theory, when the $U(1)$ symmetry
is spontaneously broken at a scale close to the string scale: the
four-dimensional gauge anomalies associated with the $U(1)$ can be
cancelled by the Green-Schwartz mechanism \cite{Green:1984sg}.  The
question of when can anomalies be cancelled by additional fermions has
not been given a general answer. It is often stated, though, that
there are cases where the anomalies can be cancelled {\it only} by the
Green-Schwartz mechanism \cite{Ibanez:1992fy, Ibanez:1994ig} (in these 
cases it is usually said that there is an ``anomalous $U(1)$'', albeit this is a
misleading phrase).

In this paper we prove that any set of fermions with arbitrary charges
under a gauge symmetry involving any number of non-Abelian and $U(1)$
groups can be embedded in an anomaly-free chiral set that contains
additional fermions---even when any ratio of charges is a rational
number.  This conclusion is far from obvious: it involves cubic
Diophantine equations ({\it i.e.}, cubic equations with integer
solutions), which include for the case of three fermions the
(in)famous Fermat's Last Theorem. 

We focus on rational charges since such charges seem more natural, but
more importantly, they solve a real problem: any $U(1)$ gauge
theory eventually hits a Landau pole unless it is embedded in a
non-Abelian group, which is possible only if these fermions 
have commensurate  charges ({\it i.e.},
rational up to a normalization of the gauge coupling) \cite{Slansky:1981yr}.
Even in string theory the gauge charges appear to be commensurate,
although we are not aware of a general proof of this statement.

In particular, we find that any ``anomalous $U(1)$'' that can be made
non-anomalous by the Green-Schwartz mechanism can also be made
non-anomalous by new fermions.  More importantly, our theorem shows that
it is possible to add new fermions such that the anomalies cancel for
any charges of the Standard Model fermions under a new $U(1)$. This is
relevant for the experimental searches for $Z^\prime$ bosons
\cite{Langacker:1984dc,*Hewett:1988xc,Carena:2004xs}, because the $Z^\prime$ couplings
to quarks and leptons are fixed, up to an overall normalization, by the
$U(1)$-charges.

In Section 2 we discuss $U(1)$ gauge anomalies, and derive our main results.
These results are then generalized to any gauge group in Sections 3 and 4.
In Section 5 we apply our results to the 
phenomenologically-interesting case of 
a $U(1)$ extension of the Standard Model gauge group.
Our conclusions are presented in Section 6.

%%%%%%%%%%%%%%%%%%%%%%%%%%%%%%%%%%%%%%%
\section{$U(1)$ gauge anomalies}
\label{sec:u1}
\setcounter{equation}{0}

Consider a set of $n$ left-handed Weyl fermions with charges $z_i$,
$i = 1, \ldots , n$ under a $U(1)$ gauge
theory. The $[U(1)]^3$ and mixed $U(1)$-gravitational
anomaly cancellation conditions are given by 
\bear 
&& \sum_{i=1}^{n} z_i^3 =0 ~,
\nonumber \\ [0.5em]
&& \sum_{i=1}^{n} z_i =0 ~.
\label{eq:u1}
\eear
We are interested in chiral sets, so the charges of the fermions must satisfy 
\be
\label{chiral}
z_i + z_j \neq 0
\ee
for any $i,j$, and in particular $z_i  \neq 0$.
For $n\le 4$, the first Eq.~(\ref{eq:u1}) can be easily solved once the constraint 
given by the second Eq.~(\ref{eq:u1}) is imposed, and the result is that 
all the fermions are vector-like, {\it i.e.}, do not satisfy  Eq.~(\ref{chiral}).
Hence, at least five chiral fermions are
needed to satisfy Eqs.~(\ref{eq:u1}) \cite{Babu:2003is,Davoudiasl:2005ks}.

When the charges are arbitrary real numbers, it is evident that 
there are solutions to Eqs.~(\ref{eq:u1}) for any $n\ge 5$. 
In realistic physics theories, however, it is generally expected that the charges 
are rational numbers up to an overall normalization of the gauge coupling. 
The reason for that is that the $U(1)$ gauge coupling 
increases with the energy, and the gauge theory appears to need a cut-off
above which some new physics would have to soften the running of the gauge 
coupling. Typically, that new physics involves the embedding of the 
$U(1)$ group 
in a non-Abelian gauge group, which guarantees that the ratio of $U(1)$ 
charges are rational numbers. We will therefore concentrate on the case
where the ratio of any two $z_i$ charges is rational. 
Furthermore, the overall normalization
of the $U(1)$ charges is arbitrary, so we can take all 
$z_i$ to be integers without
loss of generality.  For integer charges $z_i$, Eqs.~(\ref{eq:u1}) are 
equivalent to identifying the integer points in the intersection of a cubic 
hypersurface and a hyperplane in ${\mathbb R}^n$. 
The first equation (\ref{eq:u1}) is a cubic Diophantine equation,
and there are no known methods of solving it in general for a fixed 
but arbitrary value of $n$. 

%%%%%%%%%%%%%%%%%%%%%%%%%%%
\subsection{Construction of anomaly-free chiral sets}
\label{sec:proof}

There is often a more straightforward problem that arises in
model-building: given a chiral set of fermions which is anomalous,
is it possible to include more fermions such that the larger set is 
chiral and anomaly-free?  To address this issue,
we make the important observation that any fermion with integer charge
$z$ is part of the following anomaly-free set: 
\be
\label{construction}
\left\{ 1\times (z) \; , \; \; \frac{z}{6} \left(z^2-1\right) \times
(-2) \; , \; \; \frac{z}{3} \left(z^2-4\right) \times (1) \right\}~.
\ee 
where the notation $p\times (x)$ means that there are $|p|$
left-handed fermions with charge $\pm x$, the $+$ and $-$ signs
corresponding to $p > 0$ and $p<0$, respectively.  It is not surprising
that the two anomaly conditions can be satisfied by two numbers, the
numbers of fields of charge $1$ and $-2$. What is nontrivial is that
the coefficients $p$ are always integers for any
integer $z$, as a physical number of fields must be. Since fermions
with charge $\pm 1, \pm 2$ are central to this construction, we call
them basic charges for $U(1)$. If $z$ is one of the basic charges,
then the set is vector-like. Otherwise, the set is chiral.

Given a chiral set of charges $S=\{z_i, i=1,\ldots,n \}$ that may not
be anomaly-free, we construct a chiral anomaly free set that
consists only of charges in $S$ and, for each one, the appropriate
number of fermions with the basic charges, $\pm 1$ and $\pm 2$, as in
Eq.~(\ref{construction}). If some of the charges in $S$ are themselves
basic charges, then we must initially rescale all charges so this is
no longer the case. In many cases the resulting anomaly-free set 
will
still contain some vector-like pairs of $\pm 1$ and $\pm 2$, so the
final step is to remove all such pairs. This completes our
proof by construction that {\it for any chiral set of charges $S$
there is a larger set  which includes $S$ and is chiral and
anomaly-free.}

%%%%%%%%%%%%%%%%%%%%%%%%%%%
\subsection{Anomaly-free chiral sets with a small number of fermions}
\label{sec:numerical}

The above successful construction of chiral, anomaly-free sets often
requires a disturbingly large number of basic charges.
We now discuss methods for obtaining smaller sets of fermions which 
have the advantage of pushing the Landau pole to higher energies.

First, to show that smaller sets are even possible, let us first observe 
that {\it for any number $n \ge 5$ of chiral fermions there is a chiral 
set of $U(1)$ charges that is anomaly free.} To prove this statement
it is sufficient to show that there is a chiral set for each 
$n = 5,\ldots, 9$. An anomaly-free 
chiral set with arbitrary $n \ge 10$ can always be 
constructed using linear combinations of chiral sets with 
$n = 5,\ldots, 9$. In Table 1 we show anomaly-free 
chiral sets with $n = 5,\ldots, 9$ integer charges
that have the maximum charge, chosen to be positive, as small as
possible (we include all sets with the two smallest values of the 
maximum charge).

\begin{table}[t]\renewcommand{\arraystretch}{1.6}
\begin{center}\begin{tabular}{|c|c|}
\hline  number of fermions & Charges \\ \hline \hline 
5 & $ \{ 1,5,-7,-8,9 \} $ \\
  & $ \{ 2,4,-7,-9,10 \}$  \\ \hline 
6 & $ \{ 1,1,1,-4,-4,5 \}$ \\
  & $ \{-1,2,3,-5,-5,6 \}$  \\ \hline 
7 & $ \{ 1,2,2,-3,-3,-3,4 \}$  \\
  & $ \{ -1,-1,3,4,-6,-6,7 \}$  \\
  & $ \{ 1,3,-4,5,-6,-6,7 \}$  \\
  & $ \{ 2,3,3,-4,-5,-6,7 \}$  \\ \hline 
8 & $ \{ 1,1,2,3,-4,-4,-5,6 \}$ \\
  & $ \{ 2,2,2,2,-5,-5,-5,7 \}$  \\ \hline 
9 & \ $ \{ 2,2,2,-3,-3,4,-5,-5,6 \}$ \  \\
  & $ \{ 1,1,1,2,-4,5,-7,-9,10 \}$  \\ 
  & $ \{ 1,-3,4,5,5,-6,-7,-9,10 \}$  \\ \hline 
\end{tabular}
\medskip \caption{Anomaly-free 
chiral sets with $n = 5,\ldots, 9$ integer charges.}
\end{center}\end{table}

To reduce the numbers of fermions in a set $S$ we again rely on the 
construction of Eq.~(\ref{construction}). Just as the anomaly contribution 
from a single charge $z$ can be cancelled by the prescribed number of 
fields with charges $+1$ and $-2$, the reverse is also true: the anomaly 
contribution from a number of fields of charges $+1$ and $-2$ can be cancelled 
off by a single charge $z'$. In this way, large numbers of basic charges 
are exchanged for a single fermion with a large charge.

The numerical techniques that can address the problem of finding
small anomaly-free sets are defined on lattices.  For our
purposes, a lattice is any set of vectors in $\RR^n$, that is closed
under addition and subtraction ({\it i.e.}, for any two vectors $x,y$ in a
lattice, both $x+y$ and $x-y$ are also in the lattice). Each axis of
$\RR^n$ represents a possible value of fermion charge, and the
coordinates on an axis indicate the {\it number} of fermions with that
charge. Negative coordinates correspond to positive numbers of
fermions with the conjugate charge.

The set of chiral, anomaly-free sets forms a lattice, denoted $L$. For
any chiral set $X$, let $V(X)$ denote the vectorization of $X$, which
is the image of $X$ in the space $\RR^n$. For example, if $X$ is the
set $\{1,5,-7,-8,9\}$, we would define $V(X)$ to be the vector
$[1,0,0,0,1,0,-1,-1,1,0,0, \ldots]$ corresponding to the fact that
there is one fermion of charge 1, zero of charge 2, one of charge
$-7$, etc. Let $L$ denote the set of all such vectorizations $L =
\{V(X)\ |\ X$ is chiral and anomaly-free$\}$.  We can add elements of
$L$: for any two chiral, anomaly-free sets $X$ and $Y$, the sum
$V(X)+V(Y)$ corresponds to the anomaly-free chiral set that contains
all the fermions in both $X$ and $Y$, followed by the removal of all
vector-like pairs.  We can similarly subtract any two elements of $L$
to find another element of $L$; therefore $L$ is a lattice.

For a given $z$, the vectorization of the construction given in
Eq.~(\ref{construction}) is an element of $L$, which we call
$C(z)$. $C(z)$ contains one fermion with charge $z$, and the needed
number of basic charges to satisfy the anomaly equations.  The set
$\{C(z_i)\ | \ z_i \in \pm 3,4,5,\ldots\}$ actually spans $L$: any
element of $L$ can be written, by construction, as a unique linear
combination of the $C(z_i)$.

It follows that finding the smallest sets of anomaly-free chiral
fermions is the same as finding the shortest vectors in $L$. This
problem is called the ``Short Vector Problem'' and has been studied
extensively by mathematicians and computer scientists (for a 
review, see \cite{MR2120363}).  Even finding a vector which is at
most $\sqrt{2}$ times as long as the shortest vector remains an
NP-hard  problem, {\it i.e.}, at least as hard as any nondeterministic 
polynomial time problem \cite{MR1856567}. This means that for very large
numbers of fermions, it is impossible to have both accuracy and speed
in an algorithm.

To set up the problem concretely, consider searching for an anomaly-free, 
chiral set with at most $N$ fermions, whose maximum charge is $m$. A 
simple iterative approach over all possible numbers of fermions has a time 
complexity of the order of
\be
 2^{m-1} \frac{(N+m)! }{N!m!}  ~.
\ee 
Given a computing power of $10^{10}$ operations per second, it would take 
$\sim 100$ years to find the shortest solution for $N=30, m=20$. A better 
algorithm would be to search over all linear combinations of the basis 
vectors, $C(z_i)$. This has a time complexity of the order of
\be
2^{m-3} \frac{(N+m-2)!}{N! (m-2)!} ~,
\ee
and would take $\sim 1$ year for $N=30, m=20$.

Consider instead the Lenstra-Lenstra-Lovasz (LLL)
algorithm \cite{MR682664}, readily available in mathematical packages,
for attacking the shortest vector problem.
The LLL algorithm requires as input the basis 
vectors $C(z_i)$ for $|z_i|<m$, and outputs a shorter, closer to 
orthogonal set 
of basis vectors that span the same lattice.  The LLL algorithm has a time 
complexity of ${\cal O}(m^4 \log{m})$, and takes $\sim 10^{-5}$ seconds 
for $m=20$.  Note that this is polynomial in $m$ instead of exponential, 
and does not involve the number of fermions $N$ (this is possible because  
the solutions found using the LLL algorithm are by no 
means guaranteed to be minimal).  In fact, they can be up to 
$2^\frac{m-1}{2}$ times larger than the minimal solutions.  In practice, 
however, the solutions found are almost always reasonably short, and the 
significant decrease in time and ease of implementation make this approach 
worthwhile.

Since the LLL algorithm actually returns a new basis of short vectors 
which spans $L$, the algorithm can easily be adapted to solving another 
common problem in polynomial time: finding the shortest vector that 
contains a specific spectrum of fermion charges. Consider a specific set 
of charges $\{x_i\}$. To make the LLL algorithm handle this problem, 
we exchange the basis vectors $C(x_i)$ for the single basis vector $\sum_i 
C(x_i)$. The output basis set is guaranteed to include at least one short 
vector that includes the specified charges $\{x_i\}$.

Since we are interested in numbers of fermions which
are not particularly large, it may eventually prove useful to adapt even 
exponential-time solutions to the shortest vector problem in order
to identify anomaly-free sets. Although
these solutions are exponential in $m$, a recent algorithm has a time
complexity of order ${\cal O}(2^{m \log{m}})$, which take $\sim 1$
second for $m=20$ \cite{MR2120363}.

%%%%%%%%%%%%%%%%%%%%%%%%%%%%%%%%%%%%%%%%%%%%%%%%%%%%%%%%
\section{$U(1)_1\times \cdots \times U(1)_m $}
\label{sec:u12}
\setcounter{equation}{0}

Now consider a set of fermions, $\psi_i$, $i = 1,\ldots , n$, which
are charged under a $U(1)_1\times \cdots \times U(1)_m$ gauge group.
Let us denote the charges of $\psi_i$ under $U(1)_a$, $a=1,\dots, m$,
by $z_{a,i}$. The construction of anomaly-free sets will proceed as in
the case of a single $U(1)$: we will identify the number and structure of basic 
charges that are needed to cancel off the anomalies
for any single fermion. 

%%%%%%%%%%%%%%%%%%%
\subsection{$m=2$}

In the case of a $U(1)_1 \times U(1)_2$ gauge group
there are six types of anomalies: $[U(1)_1]^3$, mixed $U(1)_1$-gravitational,
$[U(1)_2]^3$, mixed $U(1)_2$-gravitational,
$[U(1)_1]^2 U(1)_2$ and $U(1)_1 [U(1)_2]^2$.
These anomalies cancel if and only if
\be
\sum_{i=1}^n z_{1,i}^3 = \sum_{i=1}^n z_{1,i} = \sum_{i=1}^n z_{2,i}^3 = \sum_{i=1}^n z_{2,i} = 
\sum_{i=1}^n z_{1,i}^2 z_{2,i} = \sum_{i=1}^n z_{1,i} z_{2,i}^2 = 0
\ee

A set of fermions is chiral with respect to $U(1)_1\times U(1)_2$ if 
\be 
z_{1, i} + z_{1, j} \neq 0  \;\;\;  {\rm or}
  \;\;\; 
z_{2, i} + z_{2, j} \neq 0 ~,
\ee
for any $i$ and $j$.
Note that a chiral set with respect to $U(1)_1\times U(1)_2$ may be 
chiral, partially vector-like, or entirely vector-like
with respect to each of the individual $U(1)$'s.

We now show that any set of fermions which is chiral with respect to $U(1)_1\times U(1)_2$
can be embedded into an anomaly-free set of chiral fermions, as we showed 
in the previous section for a $U(1)$ gauge theory.
This follows from the fact that
any fermion with integer charges $(z_{1},z_{2})$,
is part of the following anomaly-free set:
\bear
\label{anomaly-free-2} \hspace*{-3em}
\left\{ \rule{0mm}{5mm} \right. \;  (z_{1},z_{2}) \,\; , && \hspace*{-.7em}
-\frac{z_{1}z_{2}}{2} \left( z_{1} + z_{2}\right) \times ( 1,1 ) \,\; , \, \;\;\;    
-\frac{z_{1}z_{2}}{2} \left( z_{1} - z_{2}\right) \times  ( -1,1 )  \,\; , \, \;\; \;
\nonumber \\ [0.6em] 
&& \hspace*{-.7em} 
-\frac{z_{1}}{6}  \left( z_{1}^2 - 1 \right) \times ( 2,0 ) \,\; , \,\;\;  \; 
\frac{z_{1}}{3}  \left( z_{1}^2 + 3 z_{2}^2 -4 \right) \times ( 1,0 ) \,\; , \,\;\;  \;
\nonumber \\ [0.6em] 
&& \left. \hspace*{-.7em} 
-\frac{z_{2}}{6}  \left( z_{2}^2 - 1 \right) \times ( 0,2 ) \,\; , \,\;\; \;   
\frac{z_{2}}{3}  \left( 3z_{1}^2 + z_{2}^2 -4 \right) \times ( 0,1 ) \;\; \rule{0mm}{5mm} \right\} 
\eear
where the notation $p\times (x_1,x_{2})$ means that there are $|p|$ left-handed fermions 
with $U(1)_1\times U(1)_2$ charges $(x_{1},x_{2})$ for $p>0$, or 
$(-x_{1}, -x_{2})$ for $p<0$. We now have 12 basic pairs of charges  
$\pm \{(1,1), (1,-1), (2,0)$,$(1, 0), (0,2), (0,1)\}$;
that are needed to ensure anomaly cancellation. 
Note that the number of fermions with basic charges 
prescribed by Eq.~(\ref{anomaly-free-2}) is automatically an integer.
The proof for constructing 
an anomaly-free chiral set from any  chiral set $S$ proceeds 
exactly as in Section \ref{sec:proof}.

Finding small anomaly-free sets from this construction proceeds through a lattice 
construction similar to that of Section \ref{sec:numerical}. With the larger 
gauge group $U(1) \times U(1)$, the only change we make is to make each axis 
of $\RR^n$ correspond to a specific $(z_1, z_2)$ charge, instead of a single 
$U(1)$ charge $z$.  This adaptation works for finding small anomaly-free 
sets for all of the other gauge groups considered in the remainder of this paper.

%%%%%%%%%%%%%%%%%%%%%%%%%%%%%%%%%%
\subsection{$m\ge 3$}

In the case where the number of $U(1)$ gauge groups is $m \ge 3$, 
there are $m(m^2+3m+8)/6$ equations 
that must be satisfied to ensure that the theory is anomaly-free:
\be
\sum_{i=1}^n z_{a,i}=\sum_{i=1}^n z_{a,i} z_{b,i} z_{c,i}=0 ~,
\ee
for any  $a, b, c = 1, \ldots, m$.

We construct  anomaly-free chiral sets by showing that 
the anomalies of any fermion $\psi_i$ can be cancelled by the anomalies of  
a set of additional chiral fermions,
which is a generalization of the basic charges $\{(1,1)$, $(1,-1)$, $(2,0)$, $(1, 0)$, 
$(0,2)$, $(0,1)\}$ from Eq.~(\ref{anomaly-free-2}).

\begin{table}[t]\renewcommand{\arraystretch}{1.4}
\begin{center}\begin{tabular}{|c|c|c|c|}\hline
number of fermions & $U(1)_1$ & $U(1)_2$ &  $U(1)_3$ \\ \hline\hline
1              & $z_1$ & $z_2$ & $z_3$ \\ \hline
$-z_1z_2z_3$               & +1 & +1 & +1 \\ \hline
$-z_1z_2(z_1+z_2 -2z_3)/2$ & +1 & +1 & 0 \\ 
$-z_1z_2(z_1-z_2)/2$ & $-1$  & +1 & 0 \\ \hline
$-z_2z_3(z_2+z_3 -2z_1)/2$ & 0 & +1 & +1 \\ 
$-z_2z_3(z_2-z_3)/2$  & 0  & $-1$ & +1 \\ \hline
$-z_3z_1(z_3+z_1 -2z_2)/2$ & +1 & 0 & +1 \\ 
$-z_3z_1(z_3-z_1)/2$  & +1 & 0 & $-1$ \\ \hline
$-z_1 \left( z_1^2-1 \right)/6$ & +2 & 0 & 0 \\ 
$z_1 \left( z_1^2 - 4 \right)/3 + z_1(z_2^2+z_3^2-z_2z_3)$ & +1 & 0 & 0 \\ \hline
$-z_2 \left( z_2^2-1 \right)/6$ & 0 & +2 & 0  \\ 
$z_2 \left( z_2^2 - 4 \right)/3 + z_2(z_3^2+z_1^2-z_3z_1)$ & 0 & +1 & 0  \\ \hline
$-z_3 \left( z_3^2-1 \right)/6$ & 0  & 0 & +2 \\ 
$z_3 \left( z_3^2 - 4 \right)/3 + z_3(z_1^2+z_2^2-z_1z_2)$  & 0  & 0 & +1 \\ \hline
\end{tabular}
\medskip \caption{Anomaly-free chiral set of charges under three $U(1)$ groups.\label{u1n}}
\end{center}\end{table}

Consider a fermion $\psi$ with charges $(z_1, \ldots , z_m)$. Its $U(1)_a U(1)_b U(1)_c$ 
anomalies can be cancelled for any unequal $a, b, c = 1, \ldots, m$
by a number $(-z_a z_b z_c)$ 
of fermions, labelled by  $\chi_{abc}$, $a<b<c$, 
with charges $(+1,+1,+1)$ under $U(1)_a \times U(1)_b\times U(1)_c$
and charge 0 under all other groups. 
Then the  $[U(1)_a]^2 U(1)_b$ anomalies of $\psi$ and $\chi_{abc}$ 
can be cancelled for any unequal $a, b = 1, \ldots, m$
by a set of fermions composed of $N^\omega_{ab}$ fermions, labelled by $\omega_{ab}$, $a<b$, 
with charges $(+1,+1)$ under $U(1)_a\times U(1)_b$
and charge 0 under all other groups, and 
${N^\omega_{ab}}^\prime$ fermions, labelled by $\omega^\prime_{ab}$, $a<b$, 
with charges $(-1,+1)$ under $U(1)_a\times U(1)_b$
and charge 0 under all other groups, where
\bear
&& N^\omega_{ab} = z_a z_b \left[ \sum_{c=1}^m z_c - \frac{3}{2}\left(z_a+z_b\right) \right] ~,
\nonumber \\ [0.5em] 
&& {N^\omega_{ab}}^\prime = -\frac{1}{2} z_a z_b\left( z_a-z_b\right) ~.
\eear 
The remaining $[U(1)_a]^3$ and mixed $U(1)_a$-gravitational anomalies
can be cancelled for any  $a = 1, \ldots, m$ 
by a set of fermions composed of $N^\xi_a$ fermions, labelled by $\xi_a$, 
with charges $+2$ under $U(1)_a$ and charge 0 under all other groups, and 
${N^\xi_a}^\prime$ fermions, labelled by $\xi^\prime_a$, 
with charges $+1$ under $U(1)_a$ and charge 0 under all other groups, where
\bear
&& N^\xi_{a} = -\frac{1}{6} z_a \left( z_a^2-1 \right) ~,
\nonumber \\ [0.5em] 
&& {N^{\xi}_{a}}^\prime = \frac{1}{3} z_a \left( z_a^2 - 4 \right) + 
z_a\sum_{b\neq a} \left( z_b^2 - z_b \sum_{c\neq a, \, c>b} z_c \right)  ~.
\eear 

Therefore, we have constructed an anomaly-free chiral set that includes a fermion $\psi$
of arbitrary charges $(z_1, \ldots , z_m)$:
\be 
\label{anomaly-free-m} \hspace*{-.3em}
\left\{ \rule{0mm}{5mm}
\; \psi \; , \;\; 
 -z_a z_b z_c \times (\chi_{abc}) \; , \;\;  
N^\omega_{ab} \times (\omega_{ab}) \; , \;\; 
{N^\omega_{ab}}^\prime \times (\omega^\prime_{ab}) \; , \;\; 
N^\xi_{a} \times (\xi_{a}) \; , \;\;  
{N^\xi_{a}}^\prime \times (\xi^\prime_{a})   \; \rule{0mm}{5mm} \right\} 
\ee
In Table~\ref{u1n} we show the charges in the particular case $m=3$. Here, the basic 
charges are the fields $\chi, \omega, \xi$. 

The proof for constructing an anomaly-free chiral set $S'$ from any set chiral set 
$S$ proceeds exactly as in Section \ref{sec:proof}.

%%%%%%%%%%%%%%%%%%%%%%%%%%%%%%%%%%%%%%%%%%%%%%%%%%%%%
%%%%%%%%%%%%%%%%%%%%%%%%%%%%%%%%%%%%%%%%%%%%%%%%%%%%%
\section{Generalization to any gauge group}
\label{sec:sun}
\setcounter{equation}{0}

We now extend our results to $G \times U(1)$ gauge groups,
where $G$ is any non-Abelian group.
Consider a set of chiral fermions $\psi_i$, $i= 1, \ldots, n$, whose charges 
are $(R_i, z_i)$ under $G \times U(1)$. $R_i$ are some irreducible 
representations of $G$.
In addition to the $U(1)$ and $U(1)^3$ anomalies, the 
$G^3$ and $G^2 U(1)$ anomalies also must cancel 
(all other mixed anomalies are zero). 
The $G^3$ anomaly is given by 
\be
\label{G3}
A_{GGG} = \sum_i A(R_i) ~,
\ee
where the anomaly of $R_i$, $A(R_i)$, is defined by 
\be
Tr\left( \left\{T^a(R_i),T^b(R_i)\right\} T^c(R_i) \right) =  \frac{1}{2} A(R_i) d^{abc} ~.
\ee
The totally symmetric tensor $d^{abc}$ is determined by the anticommutation 
relation among the group generators $T^a(R_i)$.
The $G^2 U(1)$ anomaly is given by 
\be
\label{casimir}
A_{GG1} = \sum_i C(R_i) z_i ~,
\ee
where the Casimir of $R_i$, $C(R_i)$, is defined by 
\be
Tr\left( T^a(R_i) T^b(R_i) \right) = \delta_{ab} C(R_i) ~.
\ee
Finally, the $U(1)$ and $U(1)^3$ anomalies take the following form
up to an overall normalization:
\bear
A_{1gg} = \sum_i d(R_i) z_i ~,
\nonumber \\
A_{111} = \sum_i d(R_i) z_i^3  ~,
\eear
where $d(R_i)$ is the dimension of $R_i$.
The set of fermions $\psi_i$ is anomaly-free if 
\be
A_{GGG} = A_{GG1} = A_{1gg} = A_{111} = 0 ~.
\ee
If any of these conditions is not satisfied, then we prove by construction that 
one can add more fermions such that the larger set is both chiral
and anomaly-free.

For each fermion with charges $(R, z)$ with $z\neq 0$, we can construct an 
anomaly-free set
\be
\left\{ (R,z) \; , \; \frac{z}{6} \left(z^2-1\right) \times (R,-2)
\; , \;  \frac{z}{3} \left(z^2-4\right) \times (\overline{R},1)\; , \; 
\frac{1}{6}\left(z+1\right)\left(z+2\right)\left(z-3\right) \times (R,0) \right\} ~,  
\label{eq:sunu1}
\ee 
where $\overline{R}$ is the conjugate of $R$: $A(\overline{R}) = - A(R)$,
$C(\overline{R}) = C(R)$, $d(\overline{R}) = d(R)$.
The notation $p\times (R,x)$ means that if $p\ge 0$ then there are 
$p$  left-handed
fermions with charge $(R,x)$, while if  $p<0$ then there are 
$-p$  left-handed
fermions with charge $(\overline{R},-x)$.
The additional fermions with charge $(R,0)$ are included to make 
the entire set vector-like under the $G$ group. We have 
chosen a basis that is easy to write down explicitly, 
but in many cases is larger than necessary---one could instead make some 
of the $(R,1)$ fermions into $(\overline{R},1)$ fermions and remove 
the appropriate number of $(R,0)$ fermions 
leaving at most one fermion with charge 0 under $U(1)$.

To render the entire set \{$\psi_i$, $i= 1, \ldots, n$\} anomaly-free and chiral, 
we first rescale the $z_i$ charges to be different than $+1$ and $-2$,  
add the fermions for each field $\psi_i$ according to Eq.~(\ref{eq:sunu1}), 
then discard any remaining vector-like pairs.
Note that if any of the $z_i$ charges is zero, then one could add other 
fermions which are neutral with respect to $U(1)$ that belong to nontrivial 
representations of $G$ such that the entire set is anomaly-free and chiral 
(as done in \cite{Eichten:1982pn} for the case where $G=SU(N)$).

Typically, the total number of additional fermions can be further reduced if instead of 
fermions transforming nontrivially under $G$ we add some fermions which are 
singlets (belong to the 1 representation of $G$).
For example, consider the case where the set  \{$\psi_i$, $i= 1, \ldots, n$\} 
is anomaly-free with respect to the non-Abelian group $G$ ($A_{GGG} = 0$).
In order to cancel the  $G^2 U(1)$ anomaly we could add two more fermions with 
charges $(R,z)$ and $(\overline{R},z^\prime)$  such that 
\be
z + z^\prime = - \frac{1}{C(R)}\sum_i C(R_i) z_i ~.
\ee
One may ensure that all the U(1) charges are integers by an appropriate rescaling.
The $U(1)$ and $U(1)^3$ anomalies can be finally cancelled by including 
a number $N_1$ of fermions with charges $(1, +1)$, and 
a number $N_2$ of fermions with charges $(1, -2)$, as prescribed in 
Section 2.1:
\bear
N_1&=&\frac{1}{3} \left\{ \sum_i d(R_i) z_i \left(z_i^2-4\right)  
+ d(R) \left[ z \left(z^2-4\right)
+ z^\prime \left( {z^\prime}^2-4\right)\right]\right\} ~, \nonumber \\
N_2&=&\frac{1}{6} \left\{ \sum_i d(R_i) z_i \left(z_i^2-1\right)  
+ d(R) \left[ z \left(z^2-1\right)
+ z^\prime \left({z^\prime}^2-1\right)\right] \right\} ~.
\eear
The remarkable feature that enables this construction is that
$N_1$ and $N_2$ are integers for any integer charges $z, z^\prime, z_i$.

This procedure can immediately be extended to groups of the form 
$G_1 \times \ldots \times G_m \times  U(1)$, where $G_i$ are non-Abelian 
gauge groups.
For example, a single fermion $\psi$ with charge $(R_1, \ldots , R_m, z)$ 
is part of the anomaly free set
\bea
&& \left\{  (R_1, \ldots , R_m,  z) \; \;   , \; \; \;  \frac{z}{6} 
\left(z^2-1\right) \times (R_1,  \ldots , R_m, -2) \right.
\;  \; \; ,\\ \nonumber
&& 
 \left. \; \; \frac{z}{3} \left(z^2-4\right) \times (\overline{R}_1,\ldots, 
\overline{R}_m, 1) \; \; ,  \; \; \;
\frac{1}{6}(z+1)(z+2)(z-3) \times (R_1,  \ldots, R_m, 0) \right\} ~.  
\label{eq:sunu2}
\eea
To extend these results to arbitrary 
$G_1 \times \ldots \times G_m \times  U(1)_1\times  \ldots \times U(1)_{m^\prime}$ 
groups, one may simply use the coefficients and the set of  fermions $\chi,\omega, 
\omega^\prime, \xi, \xi^\prime$
described in Section 3.2 in place of the single $z$ charges written above.

%%%%%%%%%%%%%%%%%%%%%%%%%%%%%%%%%%%%%%%%%%%%%%%%%%%%%
%%%%%%%%%%%%%%%%%%%%%%%%%%%%%%%%%%%%%%%%%%%%%%%%%%%%%
%%%%%%%%%%%%%%%%%%%%%%%%%%%%%%%%%%%%%
\section{$U(1)$ extension of the Standard Model gauge group}
\setcounter{equation}{0}

The results presented in the previous sections have various
applications to physics beyond the Standard Model.  In this section we
study a particularly important application.  The elementary fermions
discovered in experiments so far, with charges under the
$SU(3)_C\times SU(2)_W \times U(1)_Y$ gauge group listed in
Table~\ref{TableCharge}, may be charged under a new Abelian gauge
group, $U(1)_z$, provided this is spontaneously broken.  The $U(1)_z$
charges of these Standard Model fermions determine the relative
couplings of the $Z^\prime$ boson [the heavy gauge boson associated
with $U(1)_z$], and therefore its experimental signatures. 

The discovery of a $Z^\prime$ boson with couplings to the
known fermions which are not proportional to hypercharge 
would imply the existence of certain
additional fermions \cite{Barr:1986hj} or antisymmetric tensor fields
in extra dimensions \cite{Green:1984sg}. Here, as an application
of our results, we show that {\it any}
couplings of a $Z^\prime$ boson to the Standard Model
fermions are allowed by anomaly cancellation if additional fermions
are present. For simplicity, we concentrate on generation-independent couplings.
The same method can be 
easily applied to generation-dependent couplings (in that case, though,
there are stronger phenomenological constraints from flavor-changing neutral
currents \cite{Langacker:2000ju, *Chivukula:2002ry, Carena:2004xs}).

\begin{table}[t]
\centering
\renewcommand{\arraystretch}{1.5}
\begin{tabular}{|c| |c|c|c||c|}\hline
& $SU(3)_C$ & $SU(2)_W$ & $U(1)_Y$ & $U(1)_z$ \\ \hline\hline
$q_L$ &  3 & 2 & $+1/3$ & $z_q$\\ \hline
$u_R$ &  $3$ & 1 & $4/3$ & $z_u$\\ \hline
$d_R$ &  $3$ & 1 & $-2/3$ & $z_d$\\ \hline
$l_L$ &  1 & 2 & $-1$ & $z_l$\\ \hline
$e_R$ &  1 & 1 & $-2$ & $z_e$\\ \hline
\end{tabular}

\medskip
\parbox{32em}{
\caption{Gauge charges of the Standard Model fermions in the presence of a new 
$U(1)$ group. An index labeling the three generations is implicit.
\label{TableCharge}}}
\end{table}

The $U(1)_z$ charges of the Standard Model fermions
lead in general to six different gauge anomalies:
$SU(3)_C^2 U(1)_z$, $SU(2)_W^2 U(1)_z$, $U(1)_Y^2 U(1)_z$,
$U(1)_Y U(1)_z^2$, $U(1)_z^3$ and $U(1)_z$.
It is imperative to ask whether these anomalies can all be cancelled 
simultaneously by including additional fermions. 
According to the  prescription outlined in Section
\ref{sec:sun}, one can construct anomaly-free sets for any rational 
values of the $U(1)_z$ charges.

However, realistic extensions of the Standard Model need 
additional constraints to be satisfied. One constraint 
is that there are no new stable
particles with fractional electric charges (for a review of experimental
limits, see \cite{Perl:2004qc}). 
To avoid fractional electric charges, we choose to introduce only 
fermions that transform under the Standard Model gauge group in the same 
representations (or the conjugated ones) as the observed fermions,
or are not charged under the Standard Model gauge group.
One could relax this restriction,
for example by including fermions with larger integer electric charges,
but we will not need this freedom here.

Electroweak measurements restrict severely the number of new chiral 
fermions charged under $SU(2)_W$. In order to satisfy this constraint, 
we require the new fermions to be vector-like with respect to the 
$SU(3)_C\times SU(2)_W \times U(1)_Y$, but chiral with respect to
$U(1)_z$.
Another constraint is that the both the Standard Model fermions and the new 
ones must have masses, so that some Yukawa-type couplings to Higgs fields 
need to be gauge invariant. For any specified Higgs sector, this leads 
to constraints on the $U(1)_z$ charges. However, one can keep the 
$U(1)_z$ charges of the fermions arbitrary and still give masses to all 
fermions by including a sufficient number of scalars with 
$U(1)_z$-breaking VEVs that have higher-dimensional interactions 
with the fermions. 

\begin{table}[t]
\centering
\renewcommand{\arraystretch}{1.5}
\begin{tabular}{|c| |c|c|c||c|}\hline
& $SU(3)_C$ & $SU(2)_W$ & $U(1)_Y$ & $U(1)_z$ \\ \hline\hline
$\psi_L^l$ &    &   &          &   $z_L^l$  \\ [-1em]  
           &  1 & 2 & $-1$   &         \\ [-1em]
$\psi_R^l$ &    &   &          &   $z_R^l$  \\ \hline
$\psi_L^e$ &    &   &          &   $z_L^e$  \\ [-1em]  
           &  1 & 1 & $-2$   &         \\ [-1em]
$\psi_R^e$ &    &   &          &   $z_R^e$  \\ \hline
$\psi_L^d$ &    &   &          &  $z_L^d$ \\ [-1em]  
           &  3 & 1 & $-2/3$ &          \\ [-1em]
$\psi_R^d$ &    &   &          &  $z_R^d$ \\ \hline
%%%%%%%%%%%%%
$\nu_R^j$ , $j=1, ... , N_1$ &  1 & 1 & 0 & $-1$\\ \hline
$\nu_R^{\prime k}$ , $k=1, ... , N_2$ &  1 & 1 & 0 & $+2$\\ \hline
\end{tabular}

\medskip
\parbox{32em}{
\caption{New fermions which, together with the three Standard
Model generations (see Table \ref{TableCharge}), form an anomaly-free set. 
The charges under the new $U(1)_z$ gauge group are restricted by
Eqs.~(\ref{z1-charges}) and (\ref{z2-charges}), while
$N_1$ and $N_2$ are given in Eq.~(\ref{numbers}).
\label{tab:new}}}
\end{table}

To allow for completely arbitrary charges for the
Standard Model fields under the new $U(1)_z$, the spectrum of fields
listed in Table \ref{tab:new} suffices, although other choices are
possible as long as there is at least one fermion charged under $SU(3)$,
another one charged under $SU(2)_W$ and yet another one charged under
$U(1)_Y$. 
We will eventually show that the anomaly cancellation conditions 
can be solved for arbitrary rational values of $z_q,
z_u, z_d, z_e$ and $z_l$, as well as rational values of all other charges.
Given the freedom in choosing the normalization of the gauge coupling,
we can take $z_q, z_u, z_d, z_e$ and $z_l$ to be integers.
Our method of approach is to first impose that
all of the anomalies cancel except for the $U(1)_z$-gravitational 
and $U(1)_z^3$ ones. Finding the $U(1)_z$ charges of the  $\psi$ fields
requires solving three linear equations,
corresponding to the $SU(3)_C^2 U(1)_z$,  $SU(2)_W^2 U(1)_z$ and  
$U(1)_Y^2 U(1)_z$ anomalies,
and one quadratic equation, corresponding to the $U(1)_Y U(1)_z^2$ anomaly.
Note that if we can guarantee that the $U(1)_z$ values of all fields
are still rational after having imposed these relations, then the
$U(1)_z\  \& \ U(1)_z^3$ anomaly equations can be cancelled as described
in Section \ref{sec:u1}: by adding the needed number of fields $\nu_R \ 
\& \ \nu'_R$. The remaining anomaly conditions are not affected by the
$\nu_R\  \& \ \nu'_R$ fields, so the most difficult part of satisfying the
anomaly conditions, the cubic $U(1)_z^3$ equation, is removed from the
process.

Notice that  we have chosen some of the fermions in Tables~\ref{TableCharge}
and \ref{tab:new} to be 
right-handed. Their contributions to the anomalies are the same as 
those of a left-handed fermion in the complex conjugated representation.
The three linear equations due to the $SU(3)^2 U(1)_z$, 
$SU(2)^2 U(1)_z$,  $U(1)_Y^2 U(1)_z$ anomalies constrain linear
combinations of the charges of the new fields to be 
\bea 
\label{z1-charges}
z_L^d-z_R^d &=&-3\left(2 z_q - z_u - z_d\right) ~, 
\nonumber \\ [0.5em]
z_L^l-z_R^l &=&-3\left(3 z_q + z_l\right) ~, 
\nonumber \\ [0.5em]
z_L^e-z_R^e &=&3\left(2z_q + z_u + z_e\right) ~.
\eea
To proceed with the $U(1)_Y U(1)_z^2$ anomaly cancellation condition,
given by 
\be
\label{differences}
\left(z_L^d\right)^2 - \left(z_R^d\right)^2 
+ \left(z_L^l\right)^2 - \left(z_R^l\right)^2 
+ \left(z_L^e\right)^2 - \left(z_R^e\right)^2 
= 3\left(z_q^2 - 2z_u^2+z_d^2-z_l^2+z_e^2\right)~,
\ee
we consider the particular case where 
the three remaining linear combinations of
$\psi$ charges can also be written as linear combinations of $z_q,
z_u, z_l, z_d$ and $z_e$. This reduces the $U(1)_Y U(1)_z^2$ anomaly
equation to a linear equation in the unknown coefficients which has a
three parameter solution for general values of $z_q, z_u, z_l, z_d$ and
$z_e$. We find that the charges of the new fields are given by 
\bear
\label{z2-charges}
z_L^d&=& \left(-2-\frac{3}{2} a_2 + a_1\right) z_q + \left(1+\frac{a_1}{2}\right) z_u 
+ 2 z_d - \frac{a_2}{2} z_l + \frac{a_1}{2} z_e ~,
\nonumber \\[0.5em]
z_L^l &=& \left(-6 + a_2 -a_3\right) z_q - \frac{1}{2}\left(a_2 +a_3\right)z_u - \frac{a_2}{2} z_d 
- z_l - \frac{a_3}{2} z_e ~,
\nonumber \\[0.5em]
z_L^e &=& \left(2+ a_1 - \frac{3}{2} a_3\right) z_q + \left(1-\frac{a_1}{2}\right) z_u 
- \frac{a_1}{2} z_d - \frac{a_3}{2} z_l + 2 z_e ~,
\eear
where $a_1, a_2, a_3$ are arbitrary even integers. 

To complete the proof, we add the 
necessary number of $\nu_R$ and $\nu'_R$ fields as described in Section \ref{sec:u1} 
to cancel the $U(1)_z$-gravitational and $U(1)_z^3$ anomalies:
\bear
\label{numbers}
N_1&=&\frac{1}{3}  \sum_f d_f z_f \left(z_f^2-4\right)  \nonumber \\[ 0.4em]
&=& \left(z_L^d\right)^3 - \left(z_R^d\right)^3 
+ \frac{2}{3} \left[\left(z_L^l\right)^3 - \left(z_R^l\right)^3\right]
+ \frac{1}{3} \left[\left(z_L^e\right)^3 - \left(z_R^e\right)^3\right]
\nonumber \\[0.5em] && 
+6z_q^3 - 3z_u^3-3z_d^3+2z_l^3-z_e^3 + 16z_q-4z_u~,
\nonumber \\ [1.3em]
N_2&=&\frac{1}{6} \sum_f d_f z_f \left(z_f^2-1\right)  \nonumber \\ [0.4em]
&=& \frac{1}{2}\left( N_1 -  12 z_q + 3 z_u \right) ~.
\eear
where $f$ runs over all fermions, $z_f$ is the $U(1)_z$ charge of the
fermion and $d_f$ is the dimensionality of the $SU(3)_C \times
SU(2)_W$ representation times $\pm 1$ for left-handed and right-handed
fermions, respectively.  We emphasize that Eqs.~(\ref{numbers}) yield
integer values for $N_1$ and $N_2$ for any integers $z_q, z_u, z_d,
z_e$ and $z_l$, and the values of $N_1$ and $N_2$ can be reduced using
the numerical methods of Section \ref{sec:numerical}.  Our
construction shows that all couplings of a $Z^\prime$ boson to the
Standard Model fermions are allowed by anomaly cancellation, so long
as additional fermions are present.

For illustration, let us pick some simple $U(1)_z$ charges for the 
Standard Model fermions,  $z_d=z_l=z_e=0$ and $z_q=z_u=1$, 
and compute the number of right-handed neutrinos in Table 4 
that need to be included in an anomaly free set.
We could use the freedom to choose $a_1,a_2$ and $a_3$ in order to minimize
$N_1$ and $N_2$, but for this simple case we just take
$a_1=a_2=a_3=0$.
Therefore, the charges of the $\psi$ fermions follow from Eqs.~(\ref{z2-charges}) and 
(\ref{differences}): $z_L^d = -1$, $z_R^d = 2$, $z_L^l = z_R^e = -6$, 
$z_R^l = z_L^e = 3$. Eq.~(\ref{numbers}) then gives $N_1 =-75$ and $N_2 = -42$,
which means that there are 75 right-handed neutrinos of $U(1)_z$ charge $+1$ and
42 right-handed neutrinos of $U(1)_z$ charge $-2$. This large number of 
right-handed neutrinos can be substantially reduced using  Eq.~(\ref{construction}). For example, 
the set $\{ 42\times (-2)\;,\; 75\times (+1)\}$
can be replaced by one of the following sets of five right-handed neutrinos: 
$\{ 2\times (-5)\;,\; 1 \times (-3) \;,\; 2 \times (+2)\}$ or
$\{ 1\times (-6)\;,\; 2 \times (-3) \;,\; 1 \times (+2)\;,\; 1 \times (+1) \}$.

%%%%%%%%%%%%%%%%%%%%%%%%%%%%%%%%%%%%%%%%%%%
\section{Conclusions}
\label{sec:conclusions}

The need to embed new $U(1)$ gauge groups
in non-Abelian groups forces a focus on integer-valued charges, up to
a possible rescaling of the gauge coupling constant. 
Our results show that anomaly cancellation in a gauge theory, while
highly constraining, can occur for {\it any} set of integer fermion charges
through the addition of new integer-charged fields. This is akin to
gauging $U(1)_{B-L}$ in the Standard Model: one is forced to
add a right-handed neutrino to prevent gauge anomalies from
appearing. That such anomaly-free sets exist is obvious when one
constructs vector-like sets, but highly non-trivial for chiral
integer-valued sets.

The main result is presented in Section \ref{sec:proof} for fermions
charged under a $U(1)$ gauge group, and subsequently extended to
any other gauge groups. The key observation is that there always exists a
certain {\it integer} number of basic charges that can cancel off the
anomaly from a single fermion. When the sets are large, the numerical 
techniques discussed in Section 2.2 allow a quick reduction of the set size.

Our solution is a complete description of chiral anomaly-free
sets for $U(1)^m$ gauge theories. For gauge groups that have
additional non-Abelian factors 
$G_1 \times \ldots \times G_m \times U(1)^{m^\prime}$ we
have concentrated on chiral anomaly-free sets that include
vector-like fermions with respect to some of the non-Abelian groups. 
This is sufficient to prove that any fermion can
be included in a larger chiral set of fermions that is
anomaly-free. Nevertheless, it would be interesting to extend our results and find a
complete description of anomaly-free sets under gauge groups of
the form $G_1 \times \ldots \times G_m \times U(1)^{m^\prime}$ which 
are chiral with respect to each of the $G_i$ groups.

If a gauge extension of the Standard Model is discovered, then we have argued that the
full spectrum of the new theory will still be chiral: a completely
vector-like theory would leave behind both the observed Standard Model
fermions {\it and} a set of conjugate partners after the extended
gauge symmetry breaks to $SU(3)_C \times SU(2)_W \times U(1)_Y$, which is not 
phenomenologically acceptable.
Therefore, our
results should have applications to a variety of extensions of the
Standard Model. In Section 5 we have presented a particular
application: if the Standard Model gauge group is extended to include
a new $U(1)$ group, then the Standard Model fermions may have
arbitrary rational charges under the new $U(1)$ and still the
anomalies would cancel in the presence of certain additional fermions
with rational charges.

%%%%%%%%%%%%%%%%%%%%%%%%%%%%%%%%%%%%%%%%
\bigskip\bigskip

%%%%%%%%%%%%%%%%%%%%%%%%%%%%%%%%%%%%%%%%
{\bf Acknowledgements}: We would like to thank Joe Lykken for a discussion
on commensurate charges, and Andre de Gouvea and 
Paul Langacker for comments on the manuscript. Work at ANL is supported in part by 
the US DOE, Div. of HEP, under contract W-31-109-ENG-38.
Fermilab is operated by Universities Research Association Inc.  under  
contract no. DE-AC02-76CH02000 with the DOE. 
 
\bibliography{an}
\end{document}